\documentclass{aa} 
\usepackage{graphicx}
\usepackage{txfonts}
\usepackage{natbib}
\usepackage{url}
\bibpunct{(}{)}{;}{a}{}{,} 
\newcount\longrefs
\def\aap{\ifnum\longrefs=1 {Astron.\ Astrophys.}\else 
                           {A\hbox{\rm \&}A}\fi}
\def\aapr{\ifnum\longrefs=1 {Astron.\ Astrophys.\ Rev.}\else 
                            {A\hbox{\rm \&}AR}\fi}
\def\aaps{\ifnum\longrefs=1 {Astron.\ Astrophys.\ Suppl.}\else 
                            {A\hbox{\rm \&}A Suppl.}\fi}
\def\aj{\ifnum\longrefs=1 {Astron.\ J.}\else 
                          {AJ}\fi} 
\def\ao{\ifnum\longrefs=1 {Applied Optics}\else 
                           {Appl.\ Opt.}\fi} 
\def\aspcs{\ifnum\longrefs=1 {Astron.\ Soc.\ Pacific Conf. Series}\else 
                           {ASP Conf.\ Ser.}\fi} 
\def\apj{\ifnum\longrefs=1 {Astrophys.\ J.}\else 
                           {ApJ}\fi} 
\def\apjl{\ifnum\longrefs=1 {Astrophys.\ J. Lett.}\else 
                            {ApJ}\fi} 
\def\aplett{\ifnum\longrefs=1 {Astrophys.\ J. Lett.}\else 
                            {ApJ}\fi} 
\def\apjs{\ifnum\longrefs=1 {Astrophys.\ J. Suppl.}\else 
                            {ApJS}\fi}
\def\apss{\ifnum\longrefs=1 {Astrophys.\ and Space Science}\else 
                            {Astrophys.\ Space Sci.}\fi}
\def\araa{\ifnum\longrefs=1 {Ann.\ Rev.\ Astron.\ Astrophys.}\else 
                            {ARA\hbox{\rm \&}A}\fi}
\def\azh{\ifnum\longrefs=1 {Astronomicheskii Zhurnal}\else 
                            {Astron.\ Zhur.}\fi}
\def\baas{\ifnum\longrefs=1 {Bull.\ Am.\ Astron.\ Soc.}\else 
                            {BAAS}\fi}
\def\bain{\ifnum\longrefs=1 {Bull.\ Astronom.\ Institutes Netherlands}\else
                            {Bull.\ Astr.\ Inst.\ Neth.}\fi}
\def\gca{\ifnum\longrefs=1 {Geochim.\ Cosmochim.\ Acta}\else 
                           {Geochim.\ Cosmochim.\ Acta}\fi}
\def\grl{\ifnum\longrefs=1 {Geophys.\ Res.\ Lett.}\else 
                           {Geoph.\ Res.\ Lett.}\fi}
\def\iaucirc{\ifnum\longrefs=1 {IAU Circulars}\else 
                          {IAU Circ.}\fi}
\def\ip{\ifnum\longrefs=1 {in press}\else 
                          {in press}\fi}
\def\jgr{\ifnum\longrefs=1 {J.\ Geophys.\ Res.}\else 
                           {J.\ Geophys.\ Res.}\fi}  
\def\jrasc{\ifnum\longrefs=1 {J.\ Royal Astron.\ Soc.\ Canada}\else 
                           {JRAS Can.}\fi}  
\def\mnras{\ifnum\longrefs=1 {Mon.\ Not.\ Roy.\ Astron.\ Soc.}\else 
                             {MNRAS}\fi} 
\def\nat{\ifnum\longrefs=1 {Nature}\else 
                           {Nat}\fi}
\def\pasj{\ifnum\longrefs=1 {Pub.\ Astron.\ Soc.\ Japan}\else 
                            {PASJ}\fi} 
\def\pasp{\ifnum\longrefs=1 {Pub.\ Astron.\ Soc.\ Pacific}\else 
                            {PASP}\fi} 
\def\physscr{\ifnum\longrefs=1 {Physica Scripta}\else 
                            {Phys.\ Scrip.}\fi} 
\def\planss{\ifnum\longrefs=1 {Planetary \& Space Science}\else 
                            {Plan. \& Space Sci.}\fi} 
\def\procspie{\ifnum\longrefs=1 {Proc.\ SPIE}\else 
                            {Proc.\ SPIE}\fi} 
\def\qjras{\ifnum\longrefs=1 {Quarterly J.\ Royal Astron.\ Soc.}\else 
                            {QJRAS}\fi} 
\def\sa{\ifnum\longrefs=1 {Soviet Astron..}\else 
                               {Sov.\ Astron.}\fi}
\def\skytel{\ifnum\longrefs=1 {Sky \& Telescope}\else 
                            {Sky \& Tel.}\fi} 
\def\solphys{\ifnum\longrefs=1 {Solar Phys.}\else 
                               {Sol.\ Phys.}\fi}
\def\ssr{\ifnum\longrefs=1 {Space Science Rev.}\else 
                               {Space\ Sci.\ Rev.}\fi}

\def\nl{,\ } 

\def\KIS{Kiepenheuer Institut f\"ur Sonnenphysik\nl Sch\"oneckstrasse 6\nl
         D--79104 Freiburg\nl Germany}

\def\Oslo{Institute of Theoretical Astrophysics, University of Oslo\nl 
         P.O. Box 1029, Blindern\nl N--0315 Oslo\nl Norway}

\def\SIU{Sterrekundig Instituut, Utrecht University\nl Postbus 80\,000\nl
         NL--3508 TA Utrecht\nl The Netherlands}

\def\dutch{\def\refname{Referenties}\def\abstractname{Samenvatting}%
  \def\bibname{Bibliografie}\def\chaptername{Hoofdstuk}%
  \def\appendixname{Bijlage}\def\contentsname{Inhoudsopgave}%
  \def\listfigurename{Lijst van figuren}%
  \def\listtablename{Lijst van tabellen}%
  \def\indexname{Index}\def\figurename{Figuur}\def\tablename{Tabel}%
  \def\partname{Deel}\def\enclname{Bijlage(n)}\def\ccname{Ter attentie van}%
  \def\headtoname{Aan}\def\headpagename{Pagina}%
  \def\today{\number\day\space\ifcase\month\or januari\or februari\or%
     maart\or%
     april\or mei\or juni\or juli\or augustus\or september\or oktober\or%
     november\or december\fi \space\number\year}%
  \typeout{
              >>>>> use hlatex209 for Dutch hyphenation <<<<< 
         }}

\DeclareFontFamily{OT1}{mvs}{}
\DeclareFontShape{OT1}{mvs}{m}{n}{<-> fmvr8x}{}

\newcounter{onefig} \newcounter{fignumber}
\newcount\nocaptions \newcount\nofigures \newcount\figwidth
\newcount\viewgraphs \newcount\printlabel
\long\def\skipfigure #1\viewout{}   
  \def\paper{}  \def\figlabel{} 
\long\def\nextfig#1{\setcounter{figure}{\value{fignumber}}
  \addtocounter{fignumber}{1}
  \ifnum \viewgraphs=1 \pagestyle{empty} \fi 
  \ifnum\value{onefig}=0 #1 \fi                 
  \ifnum\value{onefig}=\value{fignumber} #1 \fi}
\def\figwidths#1#2{\ifnum \nocaptions=1 #2mm \else #1mm \fi}  
\def\picplace{\framebox[80mm]{\rule{0cm}{1cm}}}
\def\paper#1{}  
\long\def\plotfig#1#2{\ifnum \nofigures=1 \picplace \else #2 \fi}
\long\def\captiontext#1{\ifnum \nofigures=1 \raggedright \fi 
   \ifnum \nocaptions=1 \paper
     \ifnum \viewgraphs=0 
       \newline  \mbox{}\hrulefill\mbox{} \newline 
       \ifnum \printlabel=1 \{{\em \figlabel}\}\newline \fi
     \fi 
   \else \ifnum \printlabel=1 \{{\em \figlabel}\}\newline \fi
     #1 \fi}

\newcount\panelwidth \newcount\panelheight 
\newcount\bxmin \newcount\bymin \newcount\bxmax \newcount\bymax
\newcount\tbxmin \newcount\tbymin
\newcount\tpanelwidth \newcount\tpanelheight \newcount\tpdif
\def\panelsize #1,#2;{\panelwidth=#1 \panelheight=#2}  
\def\setbb #1,#2;#3,#4;#5,#6;{
  \tbxmin=#1 \tbymin=#2    
  \bxmin=#3 \bymin=#4      
  \bxmax=#5 \bymax=#6}     
\def\barepanel #1{%
  \ifnum\panelheight=0 
    \tpdif=\bymax \advance\tpdif by -\bymin
    \multiply \tpdif by \panelwidth
    \tpanelheight=\tpdif
    \tpdif=\bxmax \advance\tpdif by -\bxmin
    \divide \tpanelheight by \tpdif
  \else \tpanelheight=\panelheight \fi
  \ifnum\panelwidth=0 
    \tpdif=\bxmax \advance\tpdif by -\bxmin
    \multiply \tpdif by \panelheight
    \tpanelwidth=\tpdif
    \tpdif=\bymax \advance\tpdif by -\bymin
    \divide \tpanelwidth by \tpdif
  \else \tpanelwidth=\panelwidth \fi
  \epsfig{file=#1,silent=,%
     bbllx=\bxmin bp,bblly=\bymin bp,bburx=\bxmax bp,bbury=\bymax bp,clip=,%
     width=\tpanelwidth mm,height=\tpanelheight mm}}
\def\labelypanel #1{
  \ifnum\panelheight=0 
    \tpdif=\bymax \advance\tpdif by -\bymin
    \multiply \tpdif by \panelwidth
    \tpanelheight=\tpdif
    \tpdif=\bxmax \advance\tpdif by -\bxmin
    \divide \tpanelheight by \tpdif
  \else \tpanelheight=\panelheight \fi
  \ifnum\panelwidth=0 
    \tpdif=\bxmax \advance\tpdif by -\bxmin
    \multiply \tpdif by \panelheight
    \tpanelwidth=\tpdif
    \tpdif=\bymax \advance\tpdif by -\bymin
    \divide \tpanelwidth by \tpdif
  \else \tpanelwidth=\panelwidth \fi
  \tpdif=\bxmax \advance\tpdif by -\tbxmin
  \multiply \tpanelwidth by \tpdif
  \tpdif=\bxmax \advance\tpdif by -\bxmin
  \divide \tpanelwidth by \tpdif
  \epsfig{file=#1,silent=,%
    bbllx=\tbxmin bp,bblly=\bymin bp,bburx=\bxmax bp,bbury=\bymax bp,%
    clip=,width=\tpanelwidth mm,height=\tpanelheight mm}}
\def\labelxpanel #1{%
  \ifnum\panelheight=0 
    \tpdif=\bymax \advance\tpdif by -\bymin
    \multiply \tpdif by \panelwidth
    \tpanelheight=\tpdif
    \tpdif=\bxmax \advance\tpdif by -\bxmin
    \divide \tpanelheight by \tpdif
  \else \tpanelheight=\panelheight \fi
  \ifnum\panelwidth=0 
    \tpdif=\bxmax \advance\tpdif by -\bxmin
    \multiply \tpdif by \panelheight
    \tpanelwidth=\tpdif
    \tpdif=\bymax \advance\tpdif by -\bymin
    \divide \tpanelwidth by \tpdif
  \else \tpanelwidth=\panelwidth \fi
  \tpdif=\bymax \advance\tpdif by -\tbymin
  \multiply \tpanelheight by \tpdif
  \tpdif=\bymax \advance\tpdif by -\bymin
  \divide \tpanelheight by \tpdif
  \epsfig{file=#1,silent=,%
    bbllx=\bxmin bp,bblly=\tbymin bp,bburx=\bxmax bp,bbury=\bymax bp,%
    clip=,width=\tpanelwidth mm,height=\tpanelheight mm}}
\def\labelxypanel #1{%
  \ifnum\panelheight=0 
    \tpdif=\bymax \advance\tpdif by -\bymin
    \multiply \tpdif by \panelwidth
    \tpanelheight=\tpdif
    \tpdif=\bxmax \advance\tpdif by -\bxmin
    \divide \tpanelheight by \tpdif
  \else \tpanelheight=\panelheight \fi
  \ifnum\panelwidth=0 
    \tpdif=\bxmax \advance\tpdif by -\bxmin
    \multiply \tpdif by \panelheight
    \tpanelwidth=\tpdif
    \tpdif=\bymax \advance\tpdif by -\bymin
    \divide \tpanelwidth by \tpdif
  \else \tpanelwidth=\panelwidth \fi
  \tpdif=\bxmax \advance\tpdif by -\tbxmin
  \multiply \tpanelwidth by \tpdif
  \tpdif=\bxmax \advance\tpdif by -\bxmin
  \divide \tpanelwidth by \tpdif 
  \tpdif=\bymax \advance\tpdif by -\tbymin 
  \multiply \tpanelheight by \tpdif
  \tpdif=\bymax \advance\tpdif by -\bymin
  \divide \tpanelheight by \tpdif
  \epsfig{file=#1,silent=,%
    bbllx=\tbxmin bp,bblly=\tbymin bp,bburx=\bxmax bp,bbury=\bymax bp,%
    clip=,width=\tpanelwidth mm,height=\tpanelheight mm}}

\def\CC{\par \vspace*{-2ex} \footnotesize \baselineskip=8pt \begin{verbatim}}

\long\def\startignore #1\stopignore{}   

\def\setlistparams{         
  \topsep=0.7ex                 
  \itemsep=0.7ex                
  \leftmargini=3ex}             
\setlistparams                  

\newcounter{alistindex}       

\newcounter{romenumnr}

\newlength{\minipagewidth}

\newsavebox{\boxcontent}
\newcommand{\ovalhead}[1]{
  \unitlength=1cm
  \sbox{\boxcontent}{\mbox{~~{#1}~~}}
  \begin{center}
    \ifdim\wd\boxcontent>6ex 
    \ifdim\wd\boxcontent<8cm 
    \begin{picture}(8,3) \thicklines     
      \put(4.0,0.8){\oval(8,1.6)} 
      \put(0.0,0.7){\parbox{8cm}{
         \begin{center} \usebox{\boxcontent} \end{center}}}
    \end{picture}
    \else \ifdim\wd\boxcontent<12cm 
    \begin{picture}(12,3) \thicklines     
        \put(6.0,0.8){\oval(12,1.6)} 
        \put(0.0,0.7){\parbox{12cm}{
           \begin{center} \usebox{\boxcontent} \end{center}}}
    \end{picture}
    \else
    \begin{picture}(16,3) \thicklines     
        \put(8.0,0.8){\oval(16,1.6)} 
        \put(0.0,0.7){\parbox{16cm}{
           \begin{center} \usebox{\boxcontent} \end{center}}}
    \end{picture}
    \fi \fi \fi
  \end{center}} 

\newcounter{headnr}            
\newcounter{subheadnr}[headnr]
\newcounter{subsubheadnr}[subheadnr]

\font\dropfont= cmr12 scaled \magstep5
\def\dropcap#1#2{{\noindent
    \setbox0\hbox{\dropfont #1}\setbox1\hbox{#2}\setbox2\hbox{(}%
    \count0=\ht0\advance\count0 by\dp0\count1\baselineskip
    \advance\count0 by-\ht1\advance\count0by\ht2
    \dimen1=.5ex\advance\count0by\dimen1\divide\count0 by\count1
    \advance\count0 by1\dimen0\wd0
    \advance\dimen0 by.25em\dimen1=\ht0\advance\dimen1 by-\ht1
    \global\hangindent\dimen0\global\hangafter-\count0
    \hskip-\dimen0\setbox0\hbox to\dimen0{\raise-\dimen1\box0\hss}%
    \dp0=0in\ht0=0in\box0}#2}

\def\rmit#1{{\it #1}}              
           
\def\ie{\rmit{i.e.,}}              
\def\cf{cf.}                       

\def\specchar#1{\uppercase{#1}}    

\def\CaII{\mbox{Ca\,\specchar{ii}}}

\def\level #1 #2#3#4{$#1 \: ^{#2} \mbox{#3} ^{#4}$}   

\def\rma{{\rm a}}              
\def\rmc{{\rm c}}  
\def\rmd{{\rm d}}  
\def\rme{{\rm e}} \def\rmE{{\rm E}}

\def\rmi{{\rm i}}

 \def\rmL{{\rm L}}

\def\rmo{{\rm o}}  
\def\rmp{{\rm p}}  
  
\def\rmr{{\rm r}}  
  
\def\rmt{{\rm t}} \def\rmT{{\rm T}}

\def\={\hbox{$\!=\!$}}                     

\def\mathstacksym#1#2#3#4#5{\def#1{\mathrel{\hbox to 0pt{\lower 
    #5\hbox{#3}\hss} \raise #4\hbox{#2}}}}

\mathstacksym\lta{$<$}{$\sim$}{1.5pt}{3.5pt} 
\mathstacksym\gta{$>$}{$\sim$}{1.5pt}{3.5pt} 
\mathstacksym\lrarrow{$\leftarrow$}{$\rightarrow$}{2pt}{1pt} 
\mathstacksym\lessgreat{$>$}{$<$}{3pt}{3pt} 

\def\Trad{T_{\rmr \rma \rmd}}
\def\Bnu{B_\nu} 
\def\Hii{\ion{H}{ii}}

\def\is{\ensuremath {\!=\!}}

\begin{document}

\title{Time-dependent hydrogen ionisation in 3D simulations of the
  solar chromosphere.}

\titlerunning{Time-dependent hydrogen ionisation in the solar
  chromosphere.}
  
\subtitle{I: Methods and first results}

   \author{Jorrit Leenaarts \inst{1,2}
          \and
        Sven Wedemeyer-B\"ohm \inst{3}  
   }

   \offprints{ J. Leenaarts, \\ \email{j.leenaarts@astro.uu.nl} }

   \institute{ \SIU \and \Oslo \and \KIS}
   
   \date{Received; accepted}

   \abstract{The hydrogen ionisation degree deviates substantially
     from statistical equilibrium under the conditions of the solar
     chromosphere. A realistic description of this atmospheric layer
     thus must account for time-dependent non-equilibrium effects.}
   {Advancing the realism of numerical simulations of the solar
     chromosphere by improved numerical treatment of the relevant
     physics will provide more realistic models that are essential for
     interpretation of existing and future observations.}
{An approximate method for solving the rate equations for the hydrogen
  populations was extended and implemented in the three-dimensional
  radiation (magneto-)hydrodynamics code CO5BOLD.  The method is based
  on a model atom with six energy levels and fixed radiative rates. It
  has been tested extensively in one-dimensional simulations. The
  extended method has been used to create a three-dimensional model
  that extends from the upper convection zone to the chromosphere.}
{The ionisation degree of hydrogen in our time-dependent simulation
  is comparable to the corresponding equilibrium value up to 500\,km
  above optical depth unity.  Above this height, the non-equilibrium
  ionisation degree is fairly constant over time and space, and tends
  to be at a value set by hot propagating shock waves.  The hydrogen
  level populations and electron density are much more constant than
  the corresponding values for statistical equilibrium, too.  In
  contrast, the equilibrium ionisation degree varies by more than 20
  orders of magnitude between hot, shocked regions and cool,
  non-shocked regions.  }
{The simulation shows for the first time in 3D that the chromospheric
  hydrogen ionisation degree and electron density cannot be calculated
  in equilibrium. Our simulation can provide realistic values of those 
  quantities for detailed radiative transfer computations.  }
          
   \keywords{Sun: chromosphere, radiative transfer}
  
   \maketitle

\section{Introduction}                          \label{sec:introduction}

Hydrogen is the most abundant constituent of the solar gas and the
major electron donor in the solar chromosphere. The assumption of
local thermal equilibrium (LTE) for hydrogen atoms is not valid in the
chromosphere because the radiative transitions rates dominate over the
collisional rates. Even the assumption of instantaneous statistical
equilibrium, where the level populations of each species at all times
are in equilibrium as determined by the local thermodynamic state and
the non-local radiation field, is not valid in the dynamical
chromosphere.  The timescale on which the hydrogen level populations
adjust to changes in the atmosphere (which is set by the transition
rates between levels) is too long compared to the timescale on which
the atmosphere changes.  In order to properly model this behaviour one
has to solve the rate equations for the populations of all relevant
energy levels.  We henceforth refer to this approach as time-dependent
non-LTE modelling (TD-NLTE).

  \citet{1980A&A....87..229K} 
showed that the relaxation timescale for the ionisation of hydrogen
varies from 100~s to 1000~s in the middle to upper chromosphere. 
In the nineties, time-dependent hydrogen ionisation was first
implemented in one-dimensional (1D) hydrodynamics simulations of the
solar chromosphere and corona
  \citep{1992ApJ...397L..59C, 1993ApJ...402..741H, 2002ApJ...572..626C}.
%
  \citet[][ hereafter CS2002]{2002ApJ...572..626C} 
gave a detailed analysis of dynamic hydrogen ionisation in 1D, where
they included a transition region and a corona in their model. More
recently,
  \citet{2003ApJ...589..988R} 
also implemented a time-dependent treatment of magnesium in a 1D simulation.

The next step is obviously the inclusion of time-dependent hydrogen
ionisation into 2D and 3D simulations. For reasons of numerical
stability and current computer limitations, a full time-dependent NLTE
treatment of hydrogen in the solar atmosphere is not yet possible in
two- and three-dimensional models. In the 1D case a simplified
treatment using fixed radiative rates gives a good approximation to
the full solution
 \citep{sollum1999}.
In this paper, we report on the implementation of this method in the
2D/3D radiation hydrodynamics code \mbox{CO$^5$BOLD code}
  \citep{2002AN....323..213F}. 
It yields electron densities that are more realistic than in the
original simulation, assuming LTE.  The electron densities can now
directly be used for a more realistic synthesis of chromospheric
spectral lines and continua, producing much more reliable results
compared to earlier calculations based on instantaneous statistical
equilibrium.

We give details on hydrogen ionisation in Sect.~\ref{sec:hydion} and
the numerical method in Sect.~\ref{sec:method}. The results of our
numerical simulation are described in Sect.~\ref{sec:results} and
finally discussed in Sect.~\ref{sec:discus}.

\section{Hydrogen ionisation in the solar chromosphere}
\label{sec:hydion}

In this section we recapitulate the details of chromospheric hydrogen
ionisation as found by CS2002.

Statistical equilibrium does not hold under chromospheric conditions
because the dynamical time scale of the chromosphere is much shorter
than the relaxation timescale of hydrogen level populations. The
timescale in cool inter-shock regions was found to be of the order of
$10^4$\,s. Within shocks the relaxation timescale is typically 50\,s,
owing to the higher temperatures and densities. This is still
sufficiently long to prevent equilibration of the hydrogen populations
within a shock.  Thus, the time-dependent rate equations need to be
solved to obtain the correct hydrogen level populations. Up to just
below the transition region Lyman~$\alpha$ may be regarded as being in
detailed balance for all practical purposes.  Lyman continuum
ionisation is negligible. To quote CS2002 literally: ``the dominant
hydrogen ionization process is photoionization from the second level
[\ldots]. There is a very rapid equilibration of the second and all
higher levels with the continuum, with a slow collisional leakage of
electrons from the ground state to the second level or visa versa,
depending on whether hydrogen is ionizing or recombining.''  Because
the ionisation in shocks is much faster than the post-shock
recombination, the ionisation degree and electron density tend to
represent their shock value, also in the low temperature inter-shock
regions.

\section{Method}
\label{sec:method}

\subsection{The approximate hydrogen radiative transport}
The full time-dependent NLTE radiation transport problem is as yet
unsolvable in a 3D (M-)HD code. The demands on memory and processor
time are very high, and the currently employed solution methods are
not stable enough to handle strong shock waves in 
a 3D dynamical code.  Simplifications are needed in order to compute
the non-equilibrium hydrogen ionisation.  We therefore adopt the
concept of fixed radiative rates in order to avoid computationally
expensive lambda iteration for evaluation of mean intensities.  The
new level populations can be computed from the local gas density and
temperature, the imposed radiation field, and the level populations
and electron densities from the previous time step.  We so follow the
example of
 \citet{sollum1999}
who showed that a simplified treatment of the radiative transport in
hydrogen transitions, using fixed radiative rates, gives reasonably
accurate results when compared with both static and dynamical
calculations in 1D. He compared the full and simplified treatment both
in the static case of the VAL3C atmosphere
  \citep{1981ApJS...45..635V}, 
  and in dynamical simulations with the 1D radiation hydrodynamics
  code RADYN
  \citep{1992ApJ...397L..59C}.
 He found that up to $\log{\tau_{500}}=-5.6$ the differences in the
 hydrogen level populations between the full and the simplified
 treatment were at most a few percent. Above this height (just below
 the transition region in his model) large deviations occur because of
 the neglect of Lyman transitions in the simplified treatment. Below
 we will briefly outline his method and our extensions for 2D and 3D
 simulations. See
  \citet{sollum1999}
  for more details.
  
  \subsubsection{Radiation field}

  We assume that the radiation field in each transition, both
  bound-bound and bound-free, can be described by a formal radiation
  temperature $\Trad$ so that the angle-averaged intensity $J_\nu$ at
  frequency $\nu$ is given by
  \begin{equation}
    J_\nu = 
    \frac{2 h \nu^3}{c^2} \frac{1}{\rme^{h v / k \Trad} - 1}\enspace, 
  \end{equation}
  where $h$, $c$, and $k$ are Planck's constant, the speed of light,
  and Boltzmann's constant.  The radiation temperature $\Trad$ is
  different for each transition but independent of frequency within a
  particular transition.
  However, $\Trad$ is allowed to vary over position and time. Its
  value is determined as follows: for each transition the radiation
  temperature at the upper boundary of the model atmosphere
  $\Trad^{\rmt \rmo \rmp}$ is given as input, as derived by Sollum's
  detailed 1D study. Then for each time step and each horizontal
  position we compute $z_{\rmc \rmr \rmi \rmt}(x,y)$, the smallest
  height for which
  \begin{equation}
    J_{\nu_0}(T_\rme) = 2
    J_{\nu_0}(T_{\rmr \rma \rmd}^{\rmt \rmo \rmp}).
  \end{equation}
  Here $T_\rme$ is the local gas temperature and $\nu_0$ the line
  centre frequency in the case of bound-bound transition and the
  ionisation threshold frequency in the case of bound-free
  transitions. Below this height we set $\Trad = T_\rme$, \ie\ we set
  the radiation temperature equal to the local gas temperature. Above
  this height, we set
  \begin{equation}
    J_\nu (z) = \Bnu(\Trad^{\rmt \rmo \rmp}) + 
    [ \Bnu(T_\rme(z_{\rmc \rmr \rmi \rmt})) - \Bnu(\Trad^{\rmt \rmo \rmp})] 
\left(\frac{m_{c}(z)}{m_{c}(z_{\rmc \rmr \rmi \rmt})}\right)^H.
  \end{equation}
  Here $m_{c}(z)$ is the column mass at height $z$ and $H$ is a free
    parameter that was determined by Sollum. This means we have a
  decay of $J_\nu$ as a function of column mass. From this $J_\nu(z)$
  we calculate $\Trad(z)$ for $z > z_{\rmc \rmr \rmi \rmt}$.

  This procedure ensures the radiation temperature connects smoothly
  to the gas temperature in each column at $z_{\rmc \rmr \rmi
    \rmt}$. There may be strong horizontal gradients in the radiation
  temperature around this height if the gas temperature shows strong
  horizontal gradients as well. The most important radiative
  transition for hydrogen ionisation in the chromosphere is the Balmer
  continuum. In this transition the radiation temperature ranges from
  6250~K at $z_{\rm{crit}}$ to 5237~K at the top of the computational
  domain.

  \subsubsection{Radiative bound-bound rate coefficients}

  We restate the derivation of the radiative rate coefficients
    by
  \citet{sollum1999}.

 For lines, the assumption of a narrow line ($J_\nu
  / \nu$ is constant over the line profile) gives the following
  expression for the radiative excitation rate coefficient:
  \begin{eqnarray}
    R_{lu} & = & 4 \pi \int_{0}^{\infty} \frac{\sigma_{lu}(\nu)}{h \nu} J_\nu \, \rmd \nu   \\
    & = &B_{lu} J_{\nu_o} \\ 
    & = & \frac{4 \pi^2 e^2}{h
      \nu_0 m_\rme c} f_{lu} \frac{2 h \nu_0^3}{c^2} \frac{1}{\rme^{h
        \nu_0 / k \Trad} - 1}.
  \end{eqnarray}
  Here $\sigma_{lu}(\nu)$ is the absorption cross section at frequency
  $\nu$, $B_{l u}$ is the Einstein coefficient for radiative
  excitation, $J_{\nu_0}$ is the angle-averaged radiation field at the
  line centre frequency $\nu_0$, and $f_{l u}$ is the oscillator
  strength. In addition we used the relation
\begin{equation}
  \int_{0}^{\infty} \sigma_{lu}(\nu) \, \rmd \nu = 
  \frac{h \nu_0}{4 \pi} B_{lu} = \frac{\pi e^2}{m_\rme c} f_{lu} .
\end{equation}
The bound-bound radiative deexcitation rate coefficient is
  \begin{eqnarray}
    R_{u l} & = & A_{ul} + B_{ul} J_{\nu_o} \\
    & = & \frac{g_l}{g_u} \rme^{h \nu_0 / k \Trad} R_{lu}.
  \end{eqnarray}
  where $g_l$ and $g_u$ are the statistical weights of the lower and
  upper level and $A_{ul}$ and $B_{ul}$ are the Einstein coefficients
  for spontaneous and stimulated deexcitation.

  \subsubsection{Radiative bound-free rate coefficients}
  The hydrogen bound-free excitations have a Kramers absorption cross section 
  \begin{equation}
    \sigma_{ic}(\nu) = \alpha_{_0} \left(
    \frac{\nu_0}{\nu}\right)^3, \nu > \nu_0,
  \end{equation}
  where $\alpha_0$ is the absorption cross-section at the ionisation
  edge frequency $\nu_0 = (E_{c} -E_{i})/h$, where $E_{c}$ and $E_{i}$
  are the energy of the continuum and the bound level.  In this case
  the radiative excitation rate coefficient can be written as
  \begin{eqnarray}
    R_{ic} & = & 4 \pi \int_{\nu_0}^{\infty} 
    \frac{\sigma_{ic}(\nu)}{h \nu} J_\nu \, \rmd \nu \\
    & = & \frac{8 \pi}{c^2} \alpha_0
    \nu_0^3 \int_{\nu_0}^{\infty} \frac{1}{\nu} \frac{1}{\rme^{h \nu_0
        / k \Trad} - 1} \, \rmd \nu \\ 
    & = & \frac{8 \pi}{c^2} \alpha_0
    \nu_0^3 \sum_{n=1}^\infty E_1 \left[ n \frac{h \nu_0} {k
        \Trad}\right],
  \end{eqnarray}
  where $E_1$ is the first exponential integral and $J_\nu =
  B_\nu(T_{\rmr \rma \rmd})$.  The bound-free radiative deexcitation
  rate coefficient is
  \begin{eqnarray}
    R_{ci} & = & \frac{8 \pi}{c^2} \alpha_0 \nu_0^3 \left[
      \frac{n_i}{n_c} \right]_{\rmL \rmT \rmE}
    \int_{\nu_0}^{\infty} \frac{\sigma_{ic}(\nu)}{h \nu} \left(
    \frac{2 h \nu^3}{c^2}+J_\nu \right) \rme^{-h \nu / k T_\rme} \,
    \rmd \nu .
  \end{eqnarray}
  Using the Kramers cross-section and the radiation temperature this
  can be written as
  \begin{eqnarray}
       R_{ci}  & = & \frac{8 \pi}{c^2} \alpha_0 \nu_0^3 \left[
      \frac{n_i}{n_c} \right]_{\rmL \rmT \rmE} \sum_{n=0}^\infty
    E_1 \left[ \left(n \frac{T_{\rme}}{\Trad} +1 \right) \frac{h
        \nu_0}{k T_{\rme}} \right].
  \end{eqnarray}
  Here $n_\rmi$, $n_\rmc$, and $T_{\rme}$ are the population of the
  lower level, the population of the continuum, and the local kinetic
  temperature, respectively.

\subsubsection{Collisional rate coefficients}

We take the collisional rate coefficients $C_{ij}$ from 
 \citet{1972ApJ...174..227J},
who provides approximate cross sections for collisions of hydrogen
with electrons.

\subsubsection{Electron densities}
  
As hydrogen is the most abundant element in the solar gas, ionisation
of hydrogen severely affects the electron densities, which in turn
enter into the collisional rate coefficients. LTE electron densities
are a poor choice at chromospheric heights. Hence, we adopt a hybrid
between LTE and TD-NLTE electron densities. The solution of the rate
equations in our TD-NLTE approach provides the population density of
the continuum level \Hii\ that is at the same time the contribution of
hydrogen to the electron number density.  To this we add the LTE
electron densities from the other elements by solving the Saha
equation keeping the hydrogen ionization degree constant. For
computational speed, we use a pre-calculated table that gives the
electron density as a function of hydrogen ionisation degree, gas
temperature, and mass density. Parts of the RH code
 \citep{2001ApJ...557..389U}
and partition functions by
Kurucz\footnote{\url{http://kurucz.harvard.edu/atoms/PF/}} were used
for the creation of the table.

\subsubsection{Hydrogen level populations}

The equations and methods described above are used for solving the
level population evolution equations for hydrogen,
\begin{equation}
 \frac{\partial n_i}{\partial t} + \nabla \cdot (n_i\vec{v}) =
 \sum_{j \ne i}^{n_l} n_j P_{ji} - n_i \sum_{j \ne i}^{n_l} P_{ij},
 \label{eq:RateEq} 
\end{equation}

where $n_l$,~$n_i$, and $\vec{v}$ are the total number of levels, the
number density of level~$i$, and the flow velocity. The second term on
the left hand side of Eq.~(\ref{eq:RateEq}) represents changes in the
population number due to advection by the hydrodynamic flow.  The
right hand side terms are rates into and out of level $i$ by
collisional and radiative transitions between the energy levels with
$P_{ji}$ and $P_{ij}$ being the rate coefficients for transitions
from level $j$ to $i$ and vice versa.  The rate coefficient is the
sum of collisional and radiative rate coefficients:
\begin{equation}
P_{ij} = C_{ij} + R_{ij}.
\end{equation}
%

\subsection{Numerical implementation}

Numerically, Eq.~(\ref{eq:RateEq}) is solved in two steps. In the
first step, the populations from the previous time step are advected
with the flow in the hydrodynamics computation. In the second step,
the advected populations, mass density, and temperatures are used to
calculate the new rate coefficients. Then the rate equations (without
the advection term) form a system of linear ordinary differential
equations of first order, which is solved on the time interval between
the previous and current time step using the DVODE package
  \citep{DVODE1989}. 
The solution yields the level populations for the current time
step. This system is solved with a relative accuracy of $10^{-4}$ for
each timestep. To ensure exact particle conservation we then scale the
updated population densities with the total hydrogen number density,
which is directly calculated from the mass density and the mass
fraction of hydrogen in the gas. This prevents cumulative buildup of
small errors.  The code outputs the electron densities and hydrogen
level populations for the LTE case and the TD-NLTE case for each grid
cell.


\subsubsection{The hydrodynamic simulation}

\begin{figure}
  \centering
  \includegraphics[width=\columnwidth]{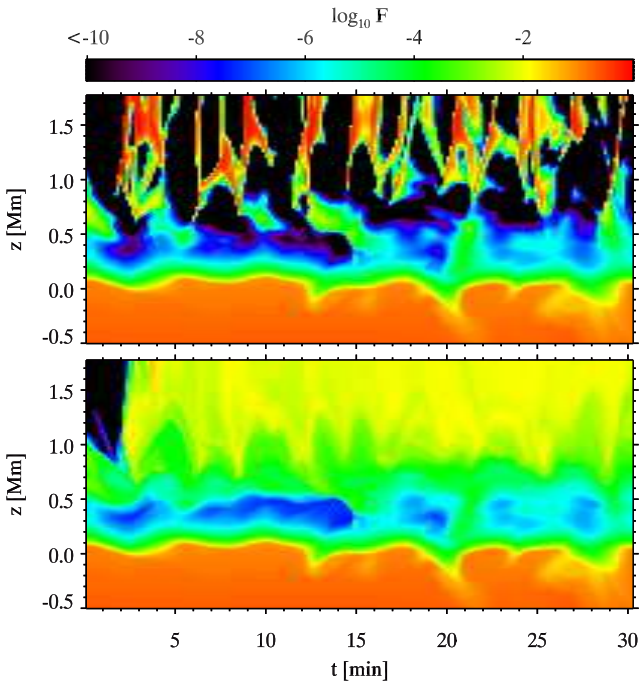}
  \caption{Time evolution of the ionisation degree $F$ in a column of the
    simulation for the first 30\,min of the simulation. Upper panel:
    LTE. Lower panel: TD-NLTE. The TD-NLTE simulation was started with
    LTE populations. The chromosphere shows large temporal variation
    in ionisation degree in the LTE case. The ionisation degree in the
    TD-NLTE case is relatively constant after the first 5\,min, when
    the first few shocks have passed.
  \label{Fig:2D_evol}}
   \end{figure}

\begin{figure*}
  \centering
  \includegraphics[width=\textwidth]{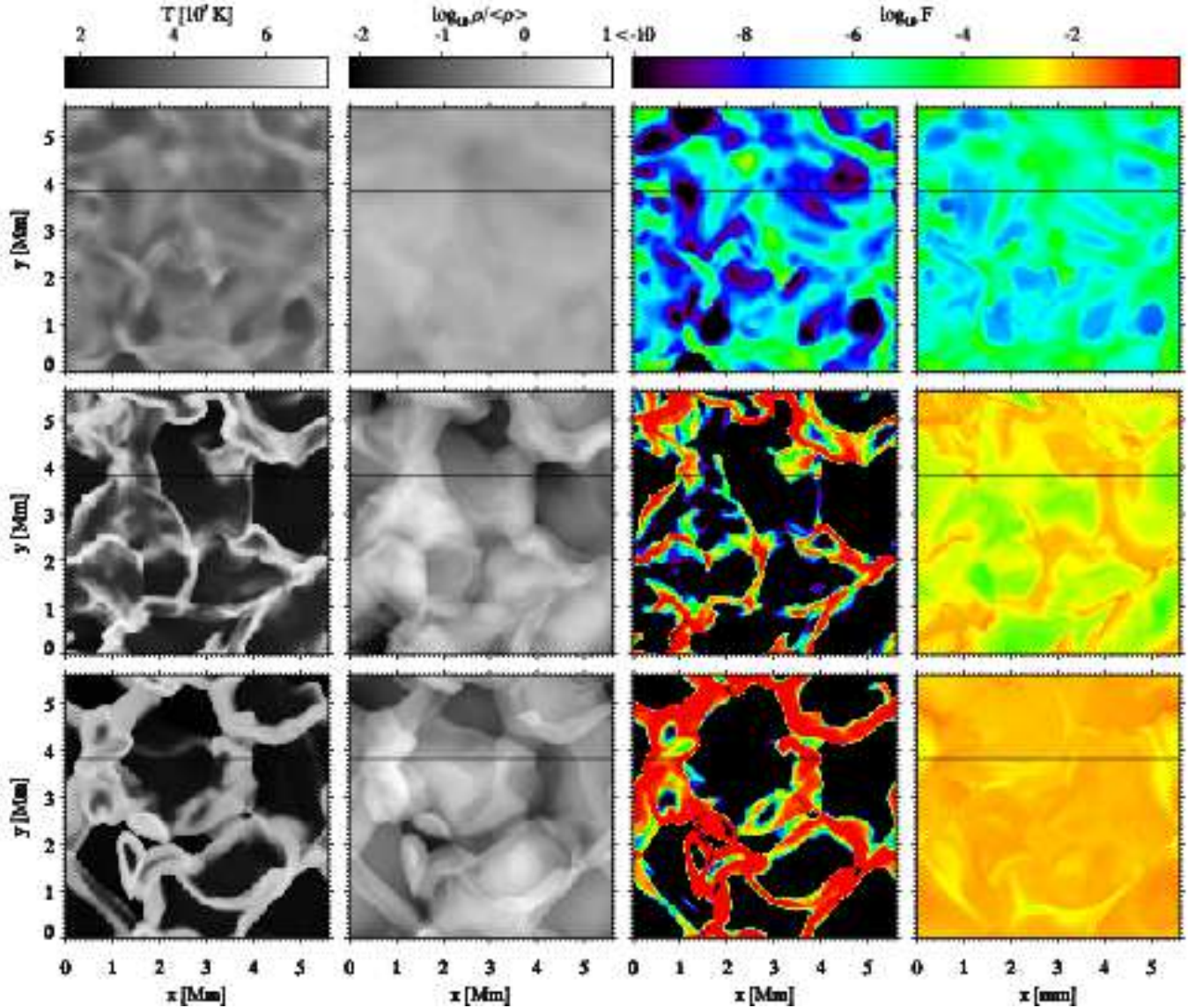}
  \caption{Horizontal slices through a simulation snapshot at $t \is
    60$\,min. Columns from left to right: gas temperature, logarithm of
    the mass density fluctuations $\log (\rho / \langle \rho \rangle)$, LTE
    ionisation degree and TD-NLTE ionisation degree $F$. 
    Top row: $z \is 0.5$~Mm;
    middle row: $z \is 1$~Mm; bottom row: $z \is 1.5$~Mm. The LTE
    and TD-NLTE ionisation degree panels have the same color scale, which
    has been clipped at $\log F \is -10$. The black line indicates the
    position of the vertical cuts of Figs.~\ref{Fig:vertslice}
    and~\ref{Fig:bslice}.
  \label{Fig:horslice}}
\end{figure*}

We performed a three-dimensional numerical simulation using an 
upgraded version of the CO$^5$BOLD code
  \citep{2002AN....323..213F}. 
The relaxed end model by
 \citet{2004A&A...414.1121W} 
was adopted as initial model.  The computational grid consists of 140
cells in each horizontal direction and 200 cells in the vertical
direction. Horizontal resolution is 40 km, vertical resolution ranges
from 46\,km at the bottom to 12\,km at the top of the computational
domain. The horizontal extent is 5600\,km. The vertical extent ranges
from -1300\,km below to 1783\,km above average Rosseland optical depth
unity. Radiative transfer is treated in the grey approximation. For
more details on the hydrodynamics code see
  \citet{wedemeyer2003} 
and
  \citet{2004A&A...414.1121W}. 
We let the simulation advance for 90\,min of solar time.  The first
30\,min are reserved to ensure relaxation of the initial conditions,
whereas the remaining 60\,min are used for detailed analysis.

\subsubsection{The model atom}

We used a five-level-plus-continuum hydrogen model atom with
collisional transitions from each level to each other level. Radiative
transitions from the ground level are not allowed, \ie\ all Lyman
transitions are put in detailed balance. Lyman cooling or heating is
thus not possible, but as our model does not include a transition zone
no significant effect even in the upper layers of our model is to be
expected. The bound-bound oscillator strengths come from
 \citet{1972ApJ...174..227J}
whereas the bound-free radiative cross sections are from
\citet{1960RPPh...23..313S}.

\section{Results} \label{sec:results}

\begin{figure}
  \centering
  \includegraphics[width=\columnwidth]{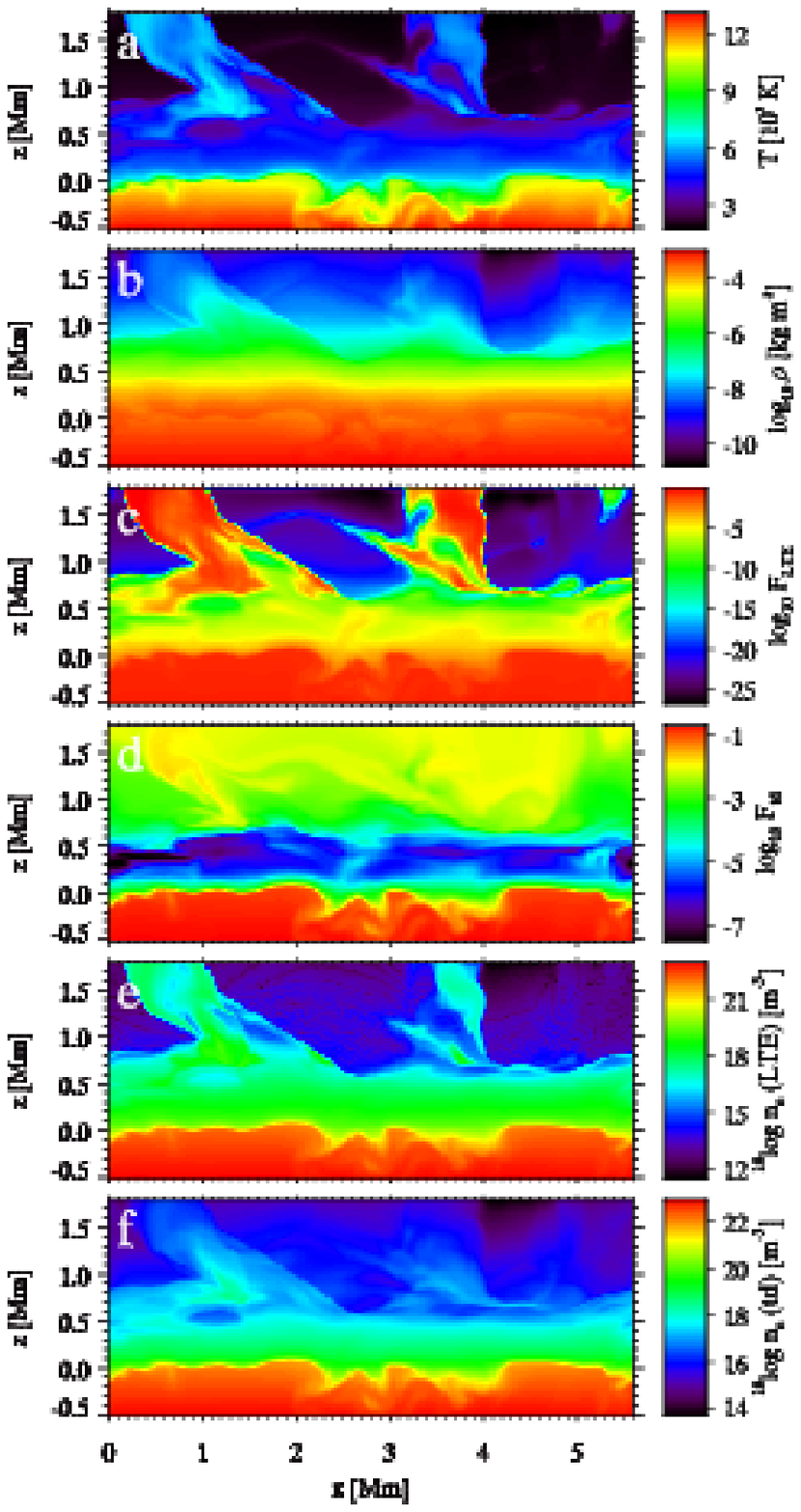}
  \caption{Vertical cut through the snapshot along the line indicated
    in Fig.~\ref{Fig:horslice}. a: gas temperature; b: mass density;
    LTE (c) and TD-NLTE (d) ionisation degree; LTE (e) and TD-NLTE (f)
    electron density.  Note that the deepest layers of the model are
    not shown.
  \label{Fig:vertslice}}
   \end{figure}

\begin{figure}
  \centering
  \includegraphics[width=\columnwidth]{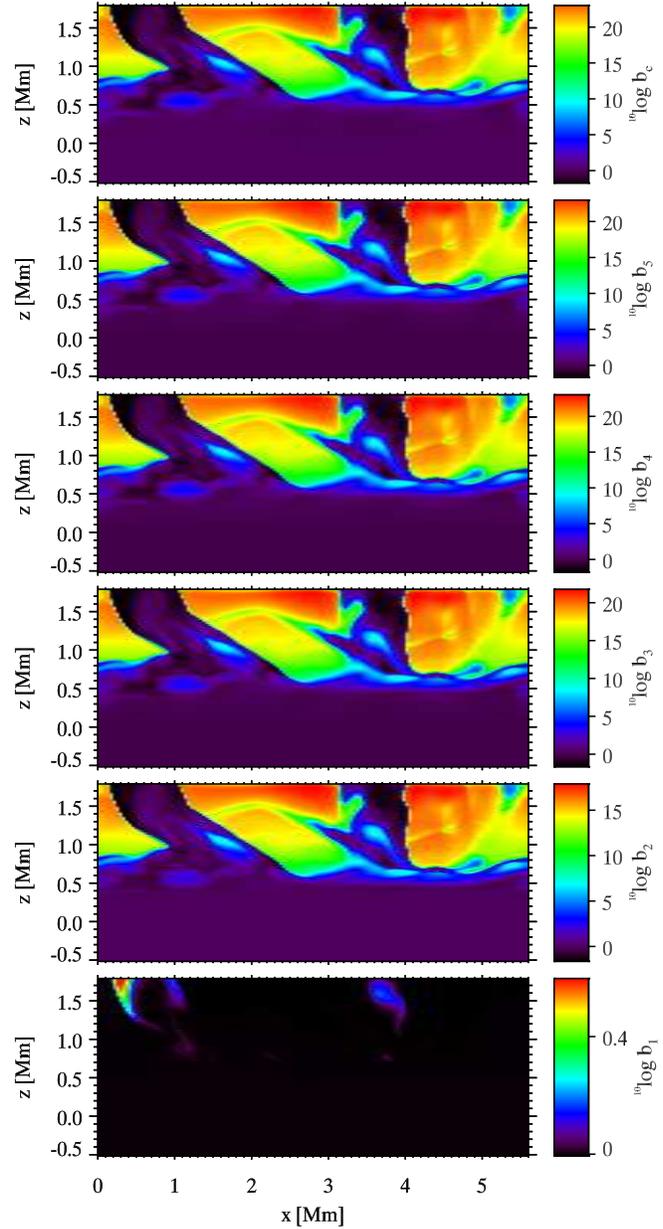}
  \caption{Departure coefficients of our model hydrogen atom in the
    same cut as Fig.~\ref{Fig:vertslice} for the continuum and level n=5
    down to level n=1 from top to bottom. The solution is nearly in
    LTE from the bottom of the computational domain up to 0.3~Mm above
    the average Rosseland optical depth unity. The largest deviations
    occur in cool chromospheric regions in between shocks, where,
    owing to the low temperature, the Saha-Boltzmann equation predicts
    a very low occupation fraction for all excited levels.
  \label{Fig:bslice}}
   \end{figure}

\begin{figure}
  \centering
  \includegraphics[width=\columnwidth]{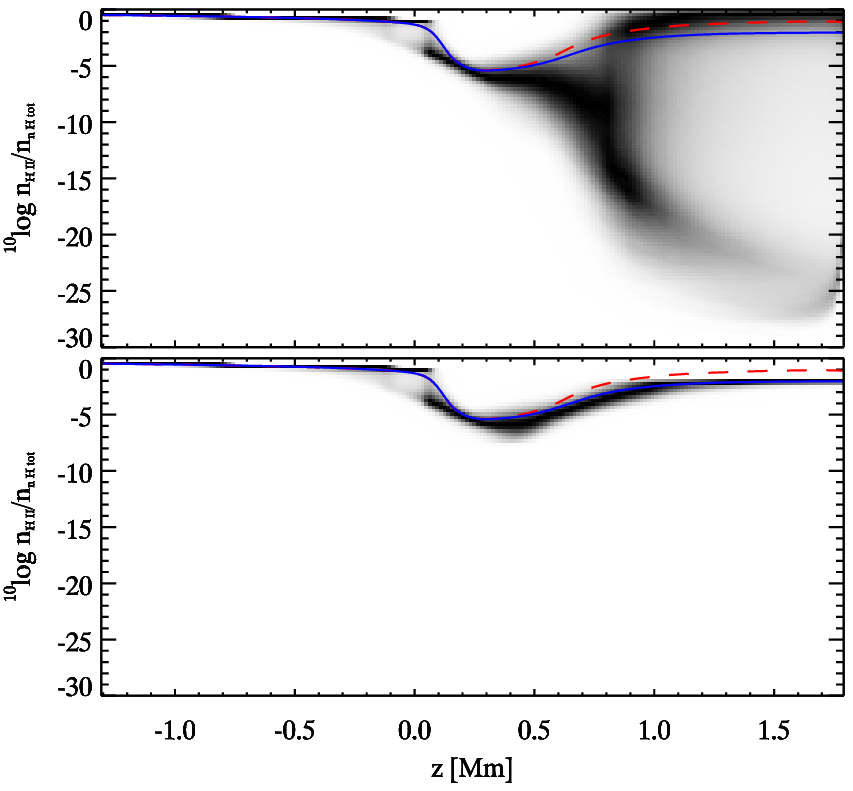}
  \caption{Histogram of the 
    logarithmic ionisation degree as a function of height for the
    LTE (upper panel) and TD-NLTE (lower panel) case. The
    averages in LTE and TD-NLTE are plotted as red dashed and blue
    solid lines, respectively. All columns have been individually
    scaled to maximum contrast to enhance visibility.
  \label{Fig:histogram}}
\end{figure}

Figure~\ref{Fig:2D_evol} shows the time evolution of the ionisation
degree $F$ along a column in the atmosphere. At $t \is 0$\,min, the
simulation is started with LTE values for the hydrogen level
populations. The LTE panel (upper) shows the chromospheric shocks as
red streaks of high ionisation in the upper part of the atmosphere,
with cool, neutral gas in between. The typical time between successive
shocks is of the order of 2--3\,min.  In contrast, the ionisation
degree in the TD-NLTE case (lower panel) reaches a dynamical
equilibrium state of fairly constant ionisation already after 5\,min,
when the first shocks have passed.

Figure~\ref{Fig:horslice} shows horizontal slices through the
simulation after 60 min. At $z \is 0.5$\,Mm the temperature and
density show little structure, as it is in between the reversed
granulation layer below
  \citep[see][]{2005A&A...431..687L}
and the onset of strong shock formation in higher layers. The
time-dependent ionisation degree is almost in LTE, showing only a
somewhat higher ionisation degree in the cooler areas, where
over-ionisation in the Balmer continuum occurs. At $z \is 1$\,Mm the
situation is different. Owing to the steady decline in average
density, the waves that are excited by the convective motions in the
photosphere have steepened into shocks. Both the temperature and
density fluctuations have increased. This has a profound effect on the
LTE ionisation degree. Because the ionisation degree is inversely
proportional to the electron density, the hydrogen ionisation at equal
temperature is much stronger than in deeper layers. In this particular
snapshot the peak ionisation degree at $z \is 1$\,Mm is 0.89 at $T \is
6813$\,K and the minimum is $3.5 \times 10^{-26}$ at $T=1871$\,K. In
the time-dependent case the ionisation degree has less extreme values,
the maximum is 0.019 at 6443\,K and the minimum is $7.8 \times
10^{-5}$ at 2495\,K.  Note also that the positions of the extrema of
the TD-NLTE ionisation degree do not at all coincide with the
positions of the LTE extrema.  The LTE ionisation degree depends on
the local mass density and temperature only, whereas in the
time-dependent case it also depends on the previous history of the
atmosphere.

At $z \is 1.5$\,Mm this effect becomes even more striking. The average
density is lower, and temperature and density fluctuations are larger,
resulting in even larger LTE ionisation degree fluctuations. In
TD-NLTE the transition rates have become so small in the cool,
non-shocked areas that there is almost no recombination between the
passages of two shock waves.  The transition rates are higher in shock
fronts as a direct result of higher temperature and density.  Although
this leads to an increased ionisation degree in shock fronts, it
remains small compared to the LTE case.  As a result the
time-dependent ionisation degree at this height is fairly constant, in
this snapshot varying only between 0.0014 and 0.021 with an average of
about 0.008.

Figure~\ref{Fig:vertslice} shows some physical quantities along a
vertical slice of the same snapshot as in Fig.~\ref{Fig:horslice}.
Panel~a shows the temperature, with granulation near 0\,Mm, reversed
granulation around 0.2\,-0.3\,Mm, and shocks above $\sim 0.7$\,Mm. The
mass density (panel~b) decreases towards larger height, while its
horizontal fluctuations increase. Panels~c and~d show the hydrogen
ionisation degree $F$ in LTE and TD-NLTE, respectively.  The LTE case
follows the increase of horizontal inhomogeneities with height, with
high ionisation in the high-temperature shock waves. In contrast, the
TD-NLTE case shows only small horizontal variation and the presence of
shocks can hardly be discerned. Finally, panels~e and~f show the
electron density in LTE and TD-NLTE. The electron density is mainly
set by the hydrogen ionisation and -- in first approximation -- can be
thought of as a multiplication of the density with $F$.  Contributions
from other elements are dominant only where $F$ is lower than
$~1\times 10^{-4}$, \ie\ around 0.4\,Mm and -- in the LTE case -- in
the cool areas in the chromosphere.  Again, in the time-dependent case
the extrema and horizontal variations are smaller in TD-NLTE than in
LTE.

Figure~\ref{Fig:bslice} shows the departure coefficients for each
level of our model atom in the same slice as Fig.~\ref{Fig:vertslice}.
The departure coefficient is defined as the ratio of the population
densities of a particular atomic energy level in NLTE and in LTE,
$b_{\rmi} = n_{\rmi} / n_\rmi^{{\rm LTE}}$.  Below $z=0.3$\,Mm the
departure coefficients for all levels are close to~1, \ie\ hydrogen is
in or very near LTE. Above these heights deviations
 occur.  The
ground level ($n =1$, lowest panel) is very close to LTE for most of
the chromosphere, except in strong shocks. In the shocks the
ionisation lags behind compared to LTE, resulting in a slight
overpopulation of the ground level (which is the reservoir where
ionised hydrogen atoms are coming from). The other levels are strongly
collisionally coupled with each other and show similar behaviour in
the chromosphere. They are slightly underpopulated in the shock area
because the ionisation time scale is comparable or slightly longer
than the typical shock crossing time. The density of the ionised state
\Hii\ (the continuum level) is correspondingly lower than in LTE. Due
to the strong coupling to the \Hii\ density, $b < 1$ for all excited
levels of neutral hydrogen ($n \is 2$ to $n \is 5$), too. In the cool
intershock areas all levels except $n \is 1$ are hugely
overpopulated. This is because the Saha and Boltzmann equations
predict extremely low population fractions for excited states and the
continuum for temperatures and electron densities typical for the
intershock regions.  Thus, these enormous overpopulations simply
illustrate the complete failure of Saha-Boltzmann equilibrium
partitioning.

Figure~\ref{Fig:histogram} shows a height-dependent histogram of the
ionisation degree. In LTE there is a strong bifurcation of the
ionisation degree as a result of the bimodal temperature distribution
in our model chromosphere
 \citep[\cf\ Fig.~7 of][]{2004A&A...414.1121W}.
In TD-NLTE this bifurcation is not present. For both cases there is a
minimum in average ionisation degree at $z \is 0.3$\,Mm, roughly
corresponding to the classical temperature minimum in 1D static
quiet-Sun models. Quite surprisingly, the average ionisation degree in
the chromosphere is lower in the TD-NLTE case than in LTE. This is
because the average is determined mainly by the high LTE ionisation
degree in the shocks whereas in the TD-NLTE case the ionisation degree
is smaller there.

\section{Discussion and conclusions}
\label{sec:discus}
\begin{figure}
  \centering
  \includegraphics[width=\columnwidth]{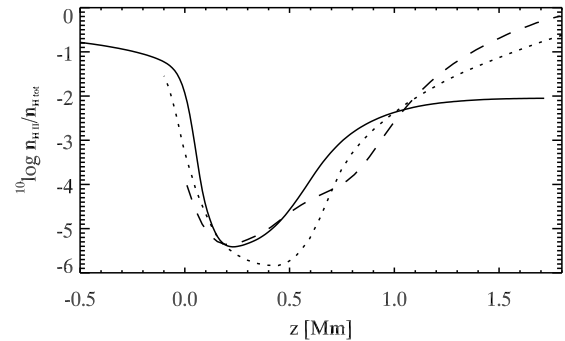}
  \caption{Average TD-NLTE ionisation degree for our CO$^5$BOLD model
    (solid), CS2002's RADYN model (dashed) and the statistical
    equilibrium FAL model C (dotted). The zero point of the
      height scale is the average $\tau_{500} \is 1$ height.
  \label{Fig:avgF}}
\end{figure}

\begin{figure}
  \centering
  \includegraphics[width=\columnwidth]{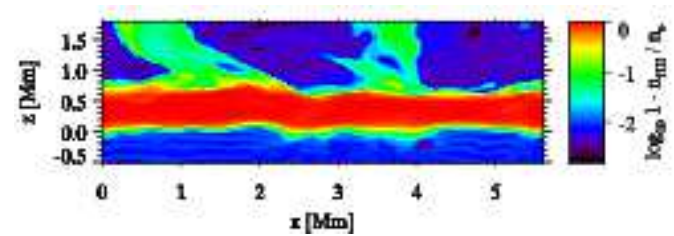}
  \caption{Vertical cut through the snapshot along the line indicated
    in Fig.~\ref{Fig:horslice} showing the relative contribution of
    other elements than hydrogen to the electron density. In the
    chromosphere the other elements contribute mostly in the high
    temperature shocks.
  \label{Fig:econt}}
\end{figure}

Figure~\ref{Fig:avgF} shows a comparison of the average ionisation
degree for our model, the detailed 1D study of CS2002, and the FAL
model C
 \citep{1993ApJ...406..319F}. 
 The latter is a static and one-dimensional semi-empirical model,
 constructed under the assumption of statistical equilibrium, while
 the RADYN and CO$^5$BOLD models are dynamic but differ in the number
 of spatial dimensions.  All models nevertheless show the same
 qualitative behaviour, with a minimum in ionisation degree between
 200 and 500\,km, and a rise of the average ionisation degree with
 height in the chromosphere. The cause of this behaviour, however, is
 different.  In the static FAL-C model, the ionisation degree reflects
 a corresponding chromospheric temperature rise.  In contrast the rise
 in ionisation degree in the dynamical RADYN and CO$^5$BOLD models is
 due to the temperature in shock fronts, which increases with height,
 without need for a rise in average temperature
 \citep{1995ApJ...440L..29C}. 

Qualitatively our CO$^5$BOLD results are comparable to the detailed 1D
study of CS2002.  We attribute the remaining differences in ionisation
degree to the different dimensionality of the code (1D vs. 3D) and the
resulting different temperature structure. 1D simulations suffer from
shock merging that lead to too high shock temperatures
 \citep{2005ApJ...631L.155U}. 
Additionally, we do not include the effect of the time-dependent
ionisation on the equation of state, causing too low shock
temperatures in our CO$^5$BOLD model 
 \citep{1992ApJ...397L..59C}.
Third, CS2002 include a transition region and corona in their model
which influence the ionisation degree in the upper chromosphere, both
by Lyman radiation ionising the upper chromosphere and the varying
height of their transition zone. Their transition zone is sometimes
considerably lower than 1.5~Mm, influencing the average ionisation
degree. Fourth, it cannot be ascertained that the fixed radiative
rates that we employ, which are accurate in 1D, are also accurate in
3D. This might affect our results as well.

Contributions to the electron density from other elements than
hydrogen are still treated under the assumption of LTE.  This will
mostly affect the upper photosphere where the hydrogen ionisation
degree is so low that metals such as iron, magnesium, and silicon are
the main electron donors.  Hydrogen is the dominant donor in all other
regions.  However, in the chromospheric shocks the other elements
contribute around 10\% of the electrons (see
Fig.~\ref{Fig:econt}). Thus we expect a small error in the
chromospheric electron density due to the LTE treatment of the other
elements (see
 \citeauthor {2003ApJ...589..988R} 
 \citeyear {2003ApJ...589..988R} 
for TD-NLTE effects on magnesium ionisation). For future simulations
that take the transition region into account, a TD-NLTE treatment of
helium might be of importance for the electron density as well, as the
higher temperatures will lead to significant helium ionisation.

All in all we conclude that our 3D simulation with time-dependent
hydrogen ionisation produces reasonably realistic -- if probably
somewhat too low -- results for the ionisation degree and electron
density, given the level of necessary simplification and resulting
increase in computational speed.

With our method we can supply snapshots of 3D (M-)HD solar atmosphere
simulations for detailed radiative transfer calculations containing
time-dependent electron densities. Up until now most MHD
simulations could only provide LTE electron densities, which are far
from realistic in the chromosphere. The TD-NLTE electron densities are
crucial in forward modelling of chromospheric diagnostics like the
\CaII~H\&K and infrared lines and (sub-)millimetre continua (see
  \citeauthor{leenaarts+wedemeyerboehm2006} 
  \citeyear{leenaarts+wedemeyerboehm2006}
 for an example of the effect on TD-NLTE electron densities on the
 latter), as collisions between electrons and atoms provide the
 coupling of the state of the gas to the radiation field via the
 source function, and affect the opacities of for example H free-free
 and H$^{-}$ bound-free and free-free transitions.


 As it is now, the code computes the ionisation degree and electron
 density from the state parameters defined by the hydrodynamics
 simulation. Future improvement of the method should take into account
 the back-coupling of the ionisation to the equation of state.  The
 deviation from instantaneous ionisation equilibrium makes it
 impossible to use pre-computed lookup tables for the equation of
 state. In addition, the lookup tables that give the bin-averaged
 source function and opacity (computed in LTE as function of
 temperature and pressure) are then no longer consistent with the
 state of the gas. New methods have to be developed to handle this
 complex problem. In particular, implementing such physics in a
 Riemann-solver as used in CO$^5$BOLD is non-trivial.  Nevertheless
 this work is currently in progress.  Significant influence on
 chromospheric wave propagation and temperature structure in the
 chromosphere can be expected
\citep[cf.][]{1992ApJ...397L..59C}.

Another necessary improvement is the relaxation of the assumption of
the fixed radiative rates. This is mainly important for the Balmer
continuum, which is the driver of the ionisation.  A radiative rate
that is proportional to the mean grey intensity, or, in the case of
multi-group opacity methods, the mean intensity in the continuum is an
obvious choice. Such a method would take the variation in the average
radiation field due to the granulation pattern into account. It will
give more accurate results in the case of strong horizontal
inhomogeneities in gas temperature and density in the upper
photosphere, as it is the case in the presence of magnetic fields.

Simulations for other stellar types than the Sun are possible but
involve re-adjustment of the radiation temperatures. This requires a
similar detailed 1D analysis and calibration as Sollum performed for
the Sun for each spectral type separately.

\begin{acknowledgements}
  The authors would like to thank O.~Steiner, J.~Bruls, and
  H.-G.~Ludwig for illuminating discussions and R.\,J.~Rutten,
  C.~Keller and the referee for improvements to the manuscript.
  H.~Uitenbroek provided his RH code.  J.~Leenaarts recognises support
  from the {\em Deutscher Akademischer Austauschdienst}, support from the
  USO-SP International Graduate School for Solar Physics (EU
  contract nr.~MEST-CT-2005-020395) and hospitality at the
  Kiepenheuer-Institut f\"ur Sonnenphysik.  S.~Wedemeyer-B\"ohm was
  supported by the {\em Deutsche Forschungs\-gemein\-schaft (DFG)},
  grant Ste~615/5.
\end{acknowledgements}

\bibliographystyle{aa} 
\bibliography{%
brown-P,%
carlsson,%
fontenla,%
freytag,%
hansteen,%
johnson,%
kneer,%
leenaarts,%
rammacher,%
seaton,%
sollum,%
stein,%
temp,%
uitenbroek,%
vernazza,%
wedemeyer,%
wedemeyer-boehm,%
wedemeyerthesis}

\end{document}